\documentclass[a4paper]{jpconf}
\begin{document}
\title{In memoriam two distinguished participants\\
       of the Bregenz Symmetries in Science Symposia:\\
       Marcos Moshinsky and Yurii Fedorovich Smirnov}

\author{Maurice R Kibler$^{a,b,c}$}

%%% NOTE FROM THE AUTHOR TO THE EDITING TEAM 
%%% VERY IMPORTANT: PLEASE, USE THE LABELLING BELOW FOR   
%%% MY AFFILIATIONS IN ORDER TO RESPECT THE RECOMMANDATIONS   
%%% OF MY SUPPORTING INSTITUTIONS 

\address{$^a$ Universit\'e de Lyon, F--69622, Lyon, France \\
         $^b$ Universit\'e Claude Bernard Lyon 1, Villeurbanne, France \\
         $^c$ CNRS/IN2P3, Institut de Physique Nucl\'eaire de Lyon, France}

%%% NOTE FROM THE AUTHOR TO THE EDITING TEAM
%%% VERY IMPORTANT: PLEASE, USE THE LABELLING ABOVE FOR   
%%% MY AFFILIATIONS IN ORDER TO RESPECT THE RECOMMANDATIONS   
%%% OF MY SUPPORTING INSTITUTIONS

\ead{m.kibler@ipnl.in2p3.fr}

\begin{abstract}
Some particular facets of the numerous works by Marcos Moshinsky and Yurii Fedorovich 
Smirnov are presented in these notes. The accent is put on some of the common interests of Yurii 
and Marcos in physics, theoretical chemistry, and mathematical physics. These notes also contain 
some more personal memories of Yurii Smirnov. 
\end{abstract}

\section{Introduction}
Yurii Fedorovich Smirnov passed away in 2008 and Marcos Moshinsky in 2009. They were 
two famous physicists with common interests in nuclear physics, atomic and molecular 
physics, and mathematical physics. More generally, both of them were at the origin of 
significant achievements in symmetry methods in physics. They actively participated in 
several Symmetries in Science Symposia in Bregenz. These two giants had parallel 
centers of interest in the sense that they developed separately some complementary 
works in nuclear physics (that led in particular to the concept of Moshinsky-Smirnov coefficients), 
dealt with some related problems in atomic, molecular and mathematical physics,  
and, finally, combined efforts to produce a beautiful and very useful book \cite{MosSmi} 
on the applications of the harmonic oscillator system to various areas of physics and chemistry. 

It is not the purpose of these notes to extensively list and analyse the numerous 
papers by Marcos and Yurii. I shall focus on some particular facets of their works. I 
had the opportunity to meet Marcos and Yurii several times in Bregenz and on several 
other occasions, and to discuss with them about Wigner-Racah algebras for finite groups, Lie groups 
and quantum groups. I had also a chance to collaborate with Yurii Smirnov. Therefore, I shall devote 
the main part of these notes to some specific domains of importance to Yurii and Marcos 
and to some more personal reminiscences on Yurii. 

\section{Marcos Moshinsky}

Marcos Moshinsky was a Mexican physicist. He was born in Kiev (Ukraine) in 1921. He
arrived as a refugee in Mexico when he was three years old and obtained 
Mexican citizenship in 1942. He received a Bachelor's degree in physics from 
the {\it Universidad Nacional Aut\'onoma de M\'exico} (U.N.A.M.) and a Ph.D. degree in 
theoretical physics, under the guidance of Eugene P. Wigner, from Princeton 
University. Marcos was also the recipient a post-doctoral fellowship at the {\it Institut Henri Poincar\'e} 
in Paris. Afterwards, he returned to Mexico and pursued a brillant career at the U.N.A.M. in Mexico City.  

Professor Marcos Moshinsky had important responsabilities as the President of 
the {\it Sociedad Mexicana de F\'\i sica}, as a member of {\it El Colegio Nacional}, 
and as a member of the editorial board of several international scientific reviews. He 
produced and/or co-produced more than 200 scientific papers and four books among which the most 
well-known are the one written in collaboration with Thomas A. Brody on transformation brackets 
for nuclear shell-model calculations \cite{BroMos} and the one with Yurii F. Smirnov on the 
applications of the harmonic oscillator in various fields of physics and quantum 
chemistry \cite{MosSmi}. He received several prizes, namely the 
{\it Premio Nacional de Ciencias y Artes} in 1968,
{\it Premio Luis Elizondo} in 1971, 
{\it Premio U.N.A.M. de Ciencias Exactas} in 1985, 
{\it Premio Pr\'\i ncipe de Asturias de Investigaci\'on Cient\'\i fica y T\'ecnica} in 1988, and
the prestigious UNESCO Science Prize in 1997 for his work in nuclear physics. He also received 
the Wigner medal in 1998. 

\section{Marcos and Yurii}

A first seminal paper by Marcos concerned the transient dynamics of particle wavefunctions, a phenomenon that 
gives rise to diffraction in time \cite{Mos1952}. However, most of his scientific work dealt with collective 
models of the nucleus, canonical transformations in quantum mechanics, and group theoretical methods in physics, 
with a special emphasis on symplectic symmetry in nuclear, atomic and molecular physics. These themes were 
of interest to Yurii too. Following the pioneering work of Talmi (who prepared his Ms.S. thesis with Guilio Racah, 
his Doctorate thesis with Wolfgang Pauli and who was a post-doctoral fellow with Eugene P. Wigner) \cite{Talmi1952}, 
both Marcos and Yurii were 
interested in the description of pairs of nucleons in a harmonic-oscillator potential. In 1959, Moshinsky developed a 
formalism to connect wavefunctions in two different coordinate systems for two particles (with identical masses) 
in a harmonic-oscillator potential \cite{Mos1959}. In this formalism, any two-particle wavefunction 
$|n_1 \ell_1, n_2 \ell_2, \lambda \mu \rangle$, expressed in coordinates with respect to the 
origin of the harmonic-oscillator potential, is a linear combination of wavefunctions 
$|n \ell, N L, \lambda \mu \rangle$, expressed in relative and centre-of-mass coordinates of the two particles. The so-called 
transformation brackets $\langle n \ell, N L, \lambda | n_1 \ell_1, n_2 \ell_2, \lambda \rangle$ make it possible to pass 
from one coordinate system to the other. Moshinsky gave an explicit expression of these coefficients in the case 
$n_1 = n_2 = 0$ and derived recurrence relations that can be used to obtain the coefficients for 
$n_1 \not= 0$ and $n_2 \not= 0$ from those for $n_1 = n_2 = 0$ \cite{Mos1959}. Along this vein, Brody and Moshinsky 
published extensive tables of transformation brackets \cite{BroMos}. At the end of the fifties, Smirnov worked out a 
similar problem, viz. the calculation of the Talmi coefficients for unequal mass nucleons, and gave solution for 
the case $n_1 \not= 0$ and $n_2 \not= 0$ \cite{Smi1961, Smi1962}. (Indeed, the transformation brackets and the Talmi 
coefficients are connected via a double Clebsch-Gordan transformation.) The coefficients 
$\langle n \ell, N L, \lambda | n_1 \ell_1, n_2 \ell_2, \lambda \rangle$, called {\it transformation brackets} by Moshinsky 
and {\it total Talmi coefficients} by Smirnov, are now referred to as Moshinsky-Smirnov coefficients. Both the 
Moshinsky-Smirnov coefficients and the Talmi coefficients were revisited at the end of the seventies in terms of 
generating functions in the framework of the approaches of Julian S. Schwinger and Valentine Bargmann to the 
harmonic-oscillator bases \cite{Mehdi1980}. (The work by Mehdi Hage Hassan \cite{Mehdi1980}, who prepared his 
State Doctorate thesis at the {\it Institut de Physique Nucl\'eaire de Lyon} and conducted his career in Beyrouth, constitutes a very deep and original 
approach to the Talmi coefficients and Moshinsky-Smirnov coefficients.) It should be noted that the transformation brackets 
or Moshinsky-Smirnov coefficients are also of importance for atoms and molecules as shown by Marcos and Yurii in their book \cite{MosSmi}
written during the time Yurii was a visiting professor at the {\it Instituto de F\'\i sica} of the 
{\it Universidad Nacional Aut\'onoma de M\'exico}. 

A second area of common interest to both Marcos and Yurii concerns the many-body 
problem considered from the point of view of unitary and symplectic groups and the use of nonlinear and 
nonbijective canonical transformations. In this vein, Moshinsky and some of his collaborators 
introduced the concept of an ambiguity group, a group required for taking into account the nonbijectivity 
of certain canonical transformations \cite{nonbijective}. Indeed, this concept is closely related to the 
one of Lie algebra under constraint \cite{KibWin}, which in turn is connected to nonbijective transformations 
like Hopf fibrations on spheres and Hopf fibrations on hyperbolids \cite{Lambert}. Among the nonbijective 
transformations, one may mention the ${\bf R}^4 \to {\bf R}^3$ Kustaanheimo-Stiefel transformation 
and the ${\bf R}^2 \to {\bf R}^2$ Levi-Civita transformation as well as their various extensions 
\cite{Lambert, MedhiKib}. In particular, the Kustaanheimo-Stiefel transformation allows one to pass from 
a four-dimensional harmonic oscillator subjected to a constraint to the three-dimensional hydrogen atom 
(see for instance \cite{KibNeg}). This subject was of interest to Yurii and he revisited the hydrogen-oscillator 
connection with Corrado Campigotto \cite{CamSmi}. 

The harmonic oscillator is a central ingredient in numerous studies 
by Smirnov and Moshinsky. Many applications of the nonrelativistic and relativistic harmonic oscillators 
to modern physics (from molecules, atoms and nuclei to quarks) were pedagogically exposed in the book by 
Marcos and Yurii \cite{MosSmi} with a special attention paid to the $n$-body problem (in the Hartree-Fock 
approximation), the nuclear collective motion and the interacting boson model. 

But their common interests were not limited to transformation brackets and harmonic oscillators. Let us briefly mention that 
both of them were also interested in group theoretical methods and symmetry methods in physics and also contributed to several 
fields of mathematical physics including, for instance, the state labelling problem, special functions, and generating functions 
(see, for example, \cite{MPSW, RRS}).

\section{Yurii Fedorovich Smirnov}

Yurii Fedorovich Smirnov was a Russian physicist. He was born in the city of Il'inskoe 
(Yaroslavl' region, Russia) in 1935. He graduated from Moscow State University. Subsequently, he 
completed his Doctorate thesis at the same university under the guidance of Yurii M. Shirokov and 
benefited from fruitful contacts with other distinguished physicists including Yakov A. Smorodinsky. He 
pursued his career in the (Skobeltsyn) Institute of Nuclear Physics and in the Physics Department of 
(Lomonosov) Moscow State University with many stays abroad. The 
last fifteen years of his life were shared between Moscow and Mexico City where he was a visiting professor and later
a professor at the {\it Instituto de F\'\i sica} and at the {\it Instituto de 
Ciencias Nucleares} of the U.N.A.M. (he spent almost 11 years in Mexico). He received prestigious awards: 
the K.D. Sinel'nikov Prize of the Ukrainian Academy of Sciences in 1982 and the M.V. Lomonosov Prize in 2002. He was 
also a member of the Academy of Sciences of Mexico. 

Yurii Smirnov authored and/or co-authored eleven books and more than 250 scientific articles. He 
also translated several scientific books into Russian. He translated, for example, a book on the 
harmonic oscillator written by Marcos Moshinsky in 1969, precisely the book that was a starting point for 
their common book on the same subject, published in 1996 \cite{MosSmi}. He was a member of the editorial 
board of several journals and a councillor of the scientific councils of the Skobeltsyn Institute of Nuclear 
Physics and of the Chemistry Department of Moscow State University, as well as of the Institute for 
Theoretical and Experimental Physics (ITEP) in Moscow. 

\section{Some personal reminiscences on Yurii} 

My first contact with the work of Yurii Smirnov dates back to 1978 when my colleague 
J. Patera showed me, on the occasion of a NATO Advanced Study Institute organised in Canada by J.C. Donini, 
a beautiful book written by D.T. Sviridov and Yu.F. Smirnov \cite{Smirnov1}. This book 
dealt with the spectroscopy of $d^N$ ions in inhomogeneous electric fields (part of a 
disciplinary domain known as crystal- and ligand-field theory in condensed matter physics and 
explored via the theory of level splitting from a theoretical point of view). In 1979, B.I. Zhilinski\u\i , 
while visiting Dijon and Lyon in France in the context of an exchange programme between 
the USSR and France, gave me another interesting book, on $f^N$ ions in 
crystalline fields, written by D.T. Sviridov, Yu.F. Smirnov and V.N. Tolstoy \cite{Smirnov2}. At 
that time, the references for mathematical aspects of crystal- and ligand-field theory were based 
on works by Y. Tanabe, S. Sugano and H. Kamimura from Japan \cite{Tanabe}, J.S. Griffith from 
England \cite{Griffith}, and Tang Au-chin and his collaborators from China \cite{Tang} (see also 
some contributions by the present author \cite{KibJMSIJQC}). The two above-mentioned books 
by Smirnov and his colleagues shed some new light on the mathematical analysis of spectroscopic 
and magnetic properties of partly filled shell ions in molecular and crystal surroundings. In 
particular, special emphasis was put on the derivation of the Wigner-Racah algebra of a finite 
group of molecular and crystallographic interest from that of the group $SO(3) \sim SU(2)/Z_2$.  

My second (indirect) contact with Yurii is related to an invitation to participate in the fifth workshop on 
{\it Symmetry Methods in Physics} in Obninsk in July 1991. Unfortunately, I did not get my visa on time 
reducing my participation to a paper in the proceedings of the workshop edited by Yu.F. Smirnov and 
R.M. Asherova \cite{KiblerObninsk}.  

In the beginning of the 1990's, I had a chance to discover another facet of Yurii's work. In 1989, 
a Russian speaking student from Switzerland, C. Campigotto, spent one year in the group of Prof.~Smirnov. He 
started working on the Kustaanheimo-Stiefel transformation, an ${\bf R}^4 \to {\bf R}^3$ 
transformation associated with the Hopf fibration ${S}^3 \to {S}^2$ with compact fiber $S^1$. (Such a transformation 
makes it possible to connect the Kepler-Coulomb system in ${\bf R}^3$ to the isotropic harmonic oscillator 
in ${\bf R}^4$.) Then, Campigotto (well-prepared by Smirnov and his team, especially Andrey M. Shirokov and Valeriy N. Tolstoy) 
came to Lyon to prepare a Doctorate thesis (partly published in \cite{Campigotto}). He defended his thesis in 1993 
with George S. Pogosyan (representing Yu.F. Smirnov) as a member of the jury. 

A fourth opportunity to work with Yurii stemmed from our mutual interest in quantum groups 
and in nuclear and atomic spectroscopy. I meet him for the first time in Dubna in 1992. We then 
started a collaboration (partly with R.M. Asherova) on $q$- and $qp$-boson calculus in the 
framework of Hopf algebras associated with the Lie algebras $su(2)$ and $su(1,1)$ \cite{KibSmiCamAsh}. In addition, 
we pursued a group-theoretical study of the Coulomb energy averaged over the $n \ell^N$--atomic 
states with a definite spin \cite{IJQC}. We also had fruitful exchanges in nuclear physics. Indeed, Prof.~Smirnov 
and his colleagues D. Bonatsos (from Greece), S.B. Drenska, P.P. Raychev and R.P. Roussev (all from Bulgaria) 
developed a model based on a one-parameter deformation of $SU(2)$ for dealing with rotational bands 
of deformed nuclei and rotational spectra of molecules \cite{Bonatsos} (see also \cite{Georgieva}). Along 
the same line, a student of mine, R. Barbier, developed in his thesis a 
two-parameter deformation of $SU(2)$ with application to superdeformed nuclei in mass region $A \sim 130-150$  
and $A \sim 190$ (partly published in \cite{Barbier}). It was a real pleasure to receive Yurii in Lyon on the 
occasion of the defence of the Barbier thesis in 1995. Indeed, from 1992 to 1995, Yurii made four stays in 
Lyon (one with his wife Rita and one with his daughter Tatyana) and we jointly participated in several meetings, 
one in Clausthal in Germany (organised by H.-D. Doebner, V.K. Dobrev and A.G. Ushveridze) and two in Bregenz in 
Austria (organised by B. Gruber and M. Ramek).

I cannot do justice to all of the fields in which 
Yurii was recognized as a superb researcher. It is enough to say that he 
contributed to many domains of mathematical physics (e.g., finite groups 
embedded in compact or locally compact groups, Lie groups and Lie algebras, 
quantum groups, special functions) and theoretical physics (e.g., nuclear, 
atomic and molecular physics, crystal- and ligand-field theory). Let me 
mention, among other fields, that he achieved alone and with collaborators 
significant advances in the theory of clustering of shell-model (nuclear) systems 
\cite{clustering}, in projection operator techniques for simple Lie groups \cite{Tolstoy}, in the 
theory of heavy ion collisions \cite{SmiTch}, and in the so-called $J$-matrix formalism for quantum 
scattering theory (see \cite{Cocoyoc} and references therein). The $J$-matrix formalism requires the 
solution of three-term recurrence relations (or second-order difference equations); 
along this line, Yurii and some of his collaborators 
published several works (see for instance \cite{Shirokov}). As another major contribution, 
at the end of the sixties he proposed in collaboration with Vladimir G. Neudatchin a method, the 
so-called (e,2e) method (an analog of the (p,2p) method used in nuclear 
physics), for the experimental investigation of the electronic structure 
of atoms, molecules and solids; this method was successfully tested in 
many laboratories around the world (see \cite{Neudatchin} and references therein).   

Yurii was also an exceptional teacher. It was very pleasant, profitable and inspiring 
to be taught by him. I personally greatly benefited from discussions with 
Yurii Smirnov.
 
\section*{Closing}

Yurii and Marcos had many students who are now famous physicists. They interacted with many 
collaborators in their countries and abroad, and had an influence on many scientists. Marcos 
Moshinsky and Yurii Fedorovich Smirnov will remain examples for many of us. We shall not 
forget them.

\end{document}